\documentclass[aps,prb,showpacs,twocolumn,amssymb]{revtex4}
\usepackage{graphicx}
\begin{document}

\draft

\title{Dynamical transport properties of NbSe$_3$ with simultaneous sliding of both charge density waves}

\author{A.~A.~Sinchenko},
\address{Kotel'nikov Institute of Radioengineering and Electronics of RAS, Mokhovaya 11-7, 125009 Moscow, Russia}
\author{P.~Monceau},
\address{Institut NEEL, CNRS and Universit\'{e} Joseph Fourier, BP 166,
38042 Grenoble, France}

\date{\today}

\begin{abstract}

Measurements of the nonlinear conductivity in NbSe$_3$ when the
high-T and the low-T charge-density waves (CDWs) are simultaneously
sliding have been performed. It is shown that the threshold electric
field, $E_{t1}$, for depinning the high-T CDW increases 4 - 5 times
at the temperature at which the low-T CDW is formed, indicating the
strong pinning effect resulting from the interaction between both
CDWs. Under application of a radio-frequency (rf) field, Shapiro
steps are never observed simultaneously for both CDWs. At the
electric field less one than for high-T CDW sliding only Shapiro
steps for low-T CDW were observed, and at higher field only Shapiro
steps for high-T CDW exist.

\end{abstract}
\pacs{71.45.Lr, 72.15.Nj, 72.70.+m,}

\maketitle

\section{ INTRODUCTION } \label{Introd}
NbSe$_3$ is an emblematic quasi-one-dimensional metal with three
pairs of metallic chains (types I, II, and III) parallel to the
monoclinic $b$ direction \cite{ECRYS11}. NbSe$_3$ undergoes two
successive Peierls transitions: at $T_{P1}=145$ K with a charge
density wave (CDW) essentially on type I chains and at $T_{P2}=59$ K
with a CDW essentially on type II chains \cite{Gruner}. The wave
vectors of both CDWs are respectively: ${\bf Q}_1=(0, 0.241, 0)$ and
${\bf Q}_2=(0.5, 0.260, 0.5)$. The Peierls transitions in this
material are not complete and ungapped carriers remain in small
pockets at the Fermi level most probably associated with type III
chains\cite{Shima,SinchenkoJPCM09}. As a result, NbSe$_3$ keeps
metallic properties down to the lowest temperatures. Application of
an electric field above a threshold value $E_{t1}$ for high-T CDW
and $E_{t2}$ for low-T CDW induces a collective electron transport
due to the coherent incommensurate CDW motion
\cite{Gorkov89,Monceau12}.

One general question immediately arises when several order
parameters coexist in the same material, such as the two CDWs in
NbSe$_3$: are they totally independent or are they interacting one
with the other? If yes, what is the result and the mechanism of this
interaction? Several attempts were already undertaken for answering
this question. X-ray scattering measurements on pinned CDWs showed
no sign of a lock-in transition \cite{Fleming83} which may be
envisioned such as:

\begin{equation}
2({\bf Q}_1 +{\bf Q}_2)\approx(1, 1, 1)
\end{equation}

From interlayer tunneling technique\cite{Orlov06} it was found that
resulting from the low-T CDW formation the CDW gap of the high-T CDW
exhibits a 10\% decrease below $T_{P2}$.

As far as the dynamical properties associated with both high-T and
low-T CDWs at temperatures $T<T_{P2}$ are concerned, contradictory
data were reported. In Refs.\onlinecite {Gill80,Richard80} it was
indicated that the depinning field, $E_{t1}$, of the high-T CDW may
saturate below $T_{P2}$ while according to
Ref.\onlinecite{Fleming80} $E_{t1}$ continuously grows without any
peculiar singularity at $T_{P2}$.

From high-resolution x-ray scattering in the presence of an applied
current below $T_{P2}$, simultaneous and oppositely directed shifts
of the relevant CDW-superlattice components along chains were
observed above a threshold current which was identified as the
depinning threshold $I_{t1}\sim 10I_{t2}$ for the more strongly
pinned high-T CDW.\cite{Ayari04} This dynamical decoupling was
explained through a sliding-induced charge transfer between the two
electronic reservoirs corresponding to the CDW wave vectors ${\bf
Q}_1$ and ${\bf Q}_2$. Using the same technique but in a different
context related to switching effects in NbSe$_3$, a dynamical
coupling was reported from analysis of the transverse structure of
both CDWs.\cite{Noh01} Note that in all these works $E_{t1}$ was not
directly determined from non-linear current-voltage characteristics
(IVs) below 60 K but estimated as being the field at which broadband
noise (BBN) increased; there is a large uncertainty in this
determination (see for instance Fig. 2 in Ref.\onlinecite{Ayari04}).

Thus, at the present time, there are no complete and reliable
measurements describing dynamical properties of NbSe$_3$ in the
temperature range corresponding to the coexistence of both CDWs. In
the following we report, for the first time, from current-voltage
characteristics (IVs), the observation of the simultaneous sliding
behavior for both CDWs. The sliding state of each CDWs is confirmed
by the observation of Shapiro steps when an radio-frequency electric
field is applied together with the dc electric field. The present
work extends the previous report\cite{Sinchenko12a} where Shapiro
steps were observed separately for each CDW.

\section{Experimental technique}
The main problem for the determination of the threshold behavior of
the high-T CDW at temperature below $T_{P2}$ from IV characteristics
is Joule heating. On the one hand, to reduce heating one needs to
use samples with high resistance and correspondingly with a small
cross-section. On the other hand,  decrease of the crystal
cross-section  leads to an exponential grow of $E_t$ because of
finite size effects.\cite{McCarten92} We have found that the
compromise between the best thermal conditions and the magnitude of
the threshold field takes place by selecting crystals with a
thickness (0.4 - 0.6) $\mu$m and for resistance at room temperature
in the range 0.5 - 2.0 k$\Omega$/mm. So, we performed our
experiments on three selected high quality NbSe$_3$ single crystals
with a thickness indicated above and a width (2 - 8)$ \mu$m. The
residual resistance ratio of the selected crystals was $R(300
K)/R(4.2 K)=(50-100)$. The crystals were cleaned in oxygen plasma
and glued on sapphire substrates by collodion. The measurements of
IV characteristics and their derivatives have been done in
conventional 4-probe configuration. Contacts were prepared from In
by cold soldering. The distance between the potential probes was 1
mm for all the samples. For studying nonstationary effects a rf
current was superposed on the dc current using the current contacts
connected to the rf generator via two capacitors.

\section{Experimental results}
In Fig.1, we have drawn the differential current-voltage
characteristics (IVs) in the temperature range 135 - 46 K for sample
No.2. The qualitatively same characteristics were observed for two
other samples. The threshold behavior corresponding to the sliding
state of the high-T CDW is clearly seen from 135 K down to 90 K: the
behavior is Ohmic for voltage less than a threshold voltage
$V_{t1}$, and for voltages in excess of this value the differential
resistance, $R_d=dV/dI$, decreases sharply. As usual $V_{t1}$
decreases from $T_{P1}$ down to 120 K and monotonically increases at
lower temperatures.

\begin{figure}[t]
\includegraphics[width=8cm]{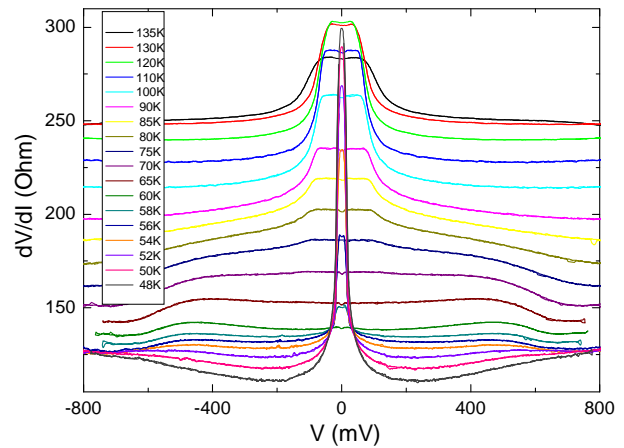}
\caption{\label{F1}(color online) Differential resistance
$R_d=dV/dI$ as a function of $V$ at different temperatures varied
from 135K to 48 K for sample No.2.}
\end{figure}

Below 90 K a new decrease of $R_d$ appears at a certain voltage
$V_{t1}^*$ that is nearly $4-5$ times larger than $V_{t1}$ and
becomes more and more pronounced in lowering  temperature. As an
example of such behavior the $dV/dI(V)$ dependence at $T=75$ K is
shown in Fig.2. When $T$ is reduced, the amplitude of the change in
$R_d$ at $V=V_{t1}$ from static to sliding decreases while that at
$V=V_{t1}^*$ increases. At $T=T_{P2}$ the first threshold at
$V=V_{t1}$ is completely indiscernible, and only the second
threshold at $V=V_{t1}^*$ remains. Below $T_{P2}$ we observe the
threshold behavior for the low-T CDW at the voltage $V_{t2}$ that is
near to $10^2$ times less than $V_{t1}^*$. Fig.3 clearly
demonstrates the singularities in the IV curve for sample No.3
corresponding to the transition into the sliding state of both CDWs
at $T=54$ K.

\begin{figure}[t]
\includegraphics[width=8cm]{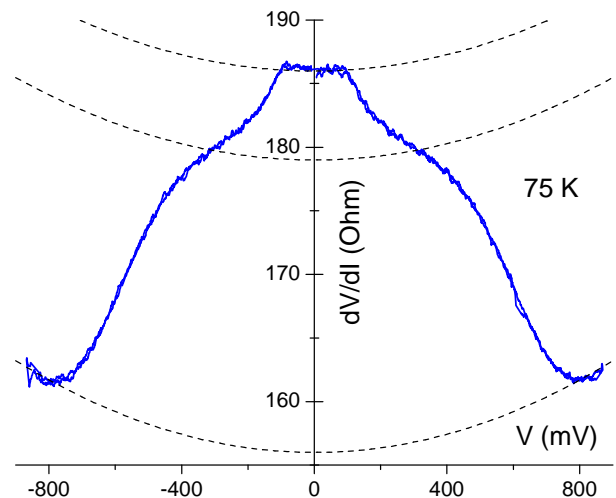}
\caption{\label{F2}(color online) $dV/dI(V)$ dependence at $T=75$ K
for sample No.2.}
\end{figure}

To prove that non-linear effects observed in dc IV curves are really
associated with CDWs sliding, we have superposed a rf current on the
dc current. It is well known that the joint application of dc and rf
driving fields leads to appearance of harmonic and sub-harmonic
Shapiro steps in the dc IV characteristics indicating sliding state
of CDW\cite{ECRYS11,Gorkov89,Monceau12}. Differential IV
characteristics for sample No.3 in the temperature range 48 - 100 K
under application of a rf field with a frequency 101 MHz are shown
in Fig.4. The Shapiro steps appear when $E>E_t$ and reflects the
synchronisation between the internal CDW sliding state and the
external frequency. As can be seen, at $T<70$ K the Shapiro steps
corresponding to the high-T CDW are observed at $E>E_{t1}^*$. Note
that in the temperature range 60 - 90 K the Shapiro steps
peculiarities are sufficiently weak compared with those at other
temperatures.

\begin{figure}[t]
\includegraphics[width=8cm]{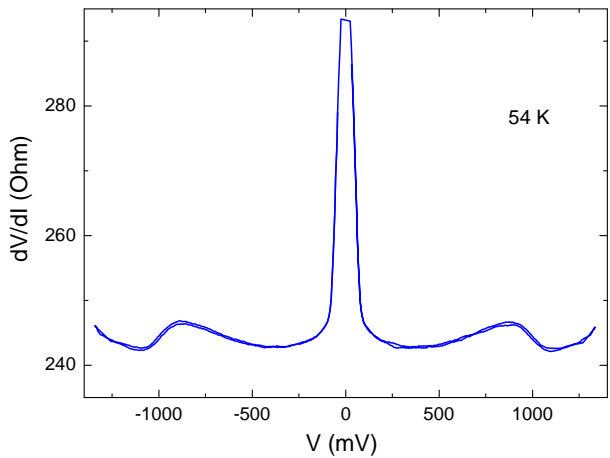}
\caption{\label{F3}(color online) $dV/dI(V)$ dependence at $T=54$ K
for sample No.3 demonstrating the simultaneous sliding of both
CDWs.}
\end{figure}

\begin{figure}[t]
\includegraphics[width=8cm]{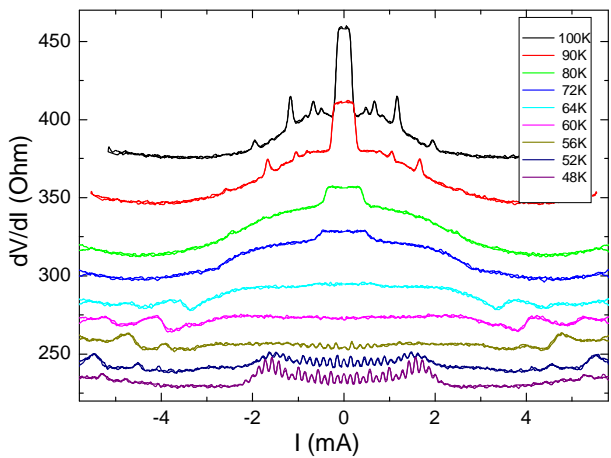}
\caption{\label{F4}(color online) $dV/dI(I)$ dependencies at
different temperatures under application of a rf field with a
frequency 101 MHz for sample No.3.}
\end{figure}

Below $T_{P2}$ we observe pronounced Shapiro steps for both CDWs as
illustrated in Fig.5 where IV curves at $T=$60, 56, 52 and 48 K are
displayed. Typically the rf amplitude was larger compared with
$V_{t2}$ and less than $V_{t1}^*$. Consequently the dc threshold at
$V_{t2}$ is totally suppressed \cite{Zettl84} while the threshold at
$V_{t1}^*$ is still observable. It is worth to note that Shapiro
steps corresponding to the low-T CDW disappear at voltages
$V>V_{t1}^*$ and only Shapiro steps associated with the high-T CDW
are observable at these voltages.

\begin{figure}[t]
\includegraphics[width=8cm]{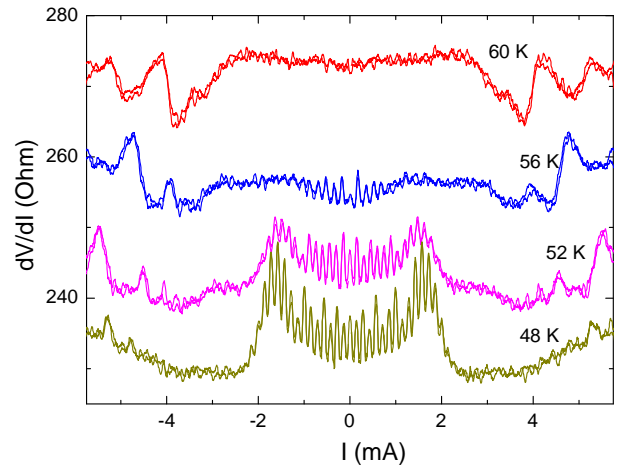}
\caption{\label{F5}(color online) The same as in Fig.\ref{F4} at
different temperatures below $T_{P2}$.}
\end{figure}

\section{Discussion}
First of all let ascertain that the drop of $R_d$ observed at
$V=V_{t1}^*$ really corresponds to the sliding of the high-T CDW. It
is well known that the fundamental frequency of the narrow-band
generation is directly proportional to the CDW
current\cite{ECRYS11,Gorkov89,Monceau12}:

\begin{equation}
\label{For1} \frac{J_{CDW}}{f_0}=c\frac{n(T)}{n(T=0)},
\end{equation}

where $n(T)$ is the number of carriers condensed into the CDW at
temperature $T$, and $c$ is a constant. As far as Shapiro steps are
concerned it is necessary to change in equation (\ref{For1})
$J_{CDW}$ to $\Delta I_{CDW}$ which corresponds to the difference of
CDW current between two neighboring harmonic Shapiro steps. $\Delta
I_{CDW}$ can be easy calculated using dc IV-characteristics. We have
calculated the relation of $\Delta I_{CDW}/f$ for the high-T CDW at
temperatures below and above $T_{P2}$ and for the low-T CDW for all
measured samples. Estimations for sample No.2 for the high-T CDW at
temperatures $T=110$ K and 54K and for the low-T CDW at $T=50$ K
with the frequency of the rf field at $f=53$ MHz are respectively
1.10; 1.17 and 1.28 $\mu$A/MHz. In accordance with
Ref.\onlinecite{Richard82}, we can derive from our experiments that
the number of condensed carriers into the CDW state is practically
equal for the low- and high-T CDW. Thus, we associate the observed
sharp drop in $R_d(V)$ dependencies at $V=V_{t1}^*$ with the
transition to the sliding state of the high-T CDW.

Using our dc curves we have also estimated the respective CDW
velocities when both CDWs are in a sliding state. Thus for sample
No.3 at $T=54$ K at the given total transport current $I=5$ mA, we
have evaluated  $I_{CDW1}\simeq0.10$ mA and $I_{CDW2}\simeq0.83$ mA.
Taking for $n=10^{21}$ cm$^-3$ (Ref.\onlinecite{Richard82}) and 2.4
$\mu$m$^2$ for the cross-section area the CDW velocities are
respectively $0.25\times10^2$ cm/s for the low-T CDW and
$2.07\times10^2$ cm/s for the high-T CDW. So, in the state when both
CDWs are in the sliding state the velocity of high-T CDW near one
order of magnitude higher than that of the low-T CDW.

In addition we can conclude that at $T<90$ K the sliding of the
high-T CDW is characterized by two threshold fields: $E_{t1}$ and
$E_{t1}^*$. The change in conductivity associated with these two
thresholds (indicated by the dashed lines in fig.2 ) is different:
when the temperature decreases, the change of conductivity below
$E_{t1}$ decreases while that below $E_{t1}^*$ increases. At
$T_{P2}=60$ K there are no more sign of the $E_{t1}$ threshold.

Fig.6 shows the temperature dependencies of threshold fields for
high- and low-T CDWs for sample No.2 in logarithmic scale. The same
characteristics were observed for other samples. The $E_{t1}(T)$
demonstrates a conventional behavior. The $E_{t1}^*$(T) decreases
with the temperature decrease from 90 down to 60 K and remains
nearly constant or demonstrates some tendency to increase at $T<60$
K. The temperature range of the observation of the coexistence
between the two thresholds $E_{t1}$ and $E_{t1}^*$ is indicated by
the dotted lines. Note, that this temperature region corresponds
well to that where low-T CDW fluctuations were
observed\cite{Brun10}. The temperature dependence of low-T threshold
field $E_{t2}(T)$ is in agreement with previous reported
data\cite{Fleming80}.

\begin{figure}[t]
\includegraphics[width=8cm]{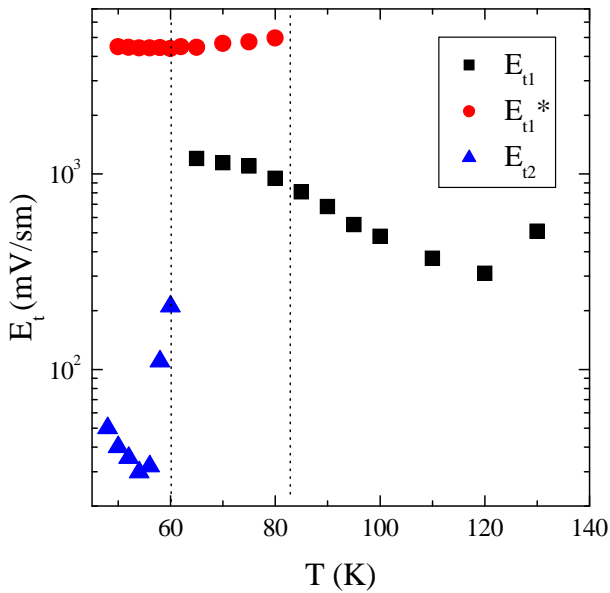}
\caption{\label{F6}(color online) Temperature dependencies the
 threshold fields $E_{t1}(T)$ (circles), $E_{t1}^*(T)$ (squares) and
$E_{t2}(T)$ (triangles) for sample No.2. Dashed lines indicate the
temperature region for fluctuations of the low-T CDW.}
\end{figure}

We can suggest the following physical picture. Above 60 K the sample
may be considered as consisting, in addition to the ordered high-T
CDW, of domains of low-T CDW fluctuations. Then the assumption is
that these fluctuations strongly pin the high-T CDW. Most probably,
the mechanism of this pinning may be a dynamical commensurability
pinning between the ordered high-T CDW and fluctuations of low-T
CDW.

As a result, in the temperature range between 60 and90 K the sliding
of the high-T CDW exhibits two threshold fields reflecting the
mixing of two types of domains. With the temperature growing the
size of fluctuation regions decreases and becomes negligibly small
at $T>90$ K where only the phase with $E_{t1}$ has been observed.
Inversely, at $T<60$ K only the phase with $E_{t1}^*$ exists. At
temperatures $75<T<85$ K both phases are in the near equal
proportion; as a result the Shapiro steps are essentially smeared at
this temperature range.

Another interesting observation of the present work is the unusual
structure of Shapiro steps when both CDWs exist simultaneously. As
it was mentioned above, in such a condition Shapiro steps
corresponding to the low-T CDW disappear at voltages $V>V_{t1}^*$
and only Shapiro steps associated with the high-T CDW are observable
at these voltages (Fig.\ref{F5}). Formally it may mean that the
motion of the high-T CDW suppresses the narrow-band-noise generation
from the low-T CDW. The complete explanation of this effect is
missing now.

Recently, a new view on the origin of the narrow band noise
generation in CDW compounds was suggested\cite{Sinchenko12a}, namely
that Shapiro steps in CDW compounds are the result of electronic
transport transversely to CDW chains. Such assumption is based on
theory proposed in Refs.\onlinecite{ArtemenkoVolkov84,Artemenko97}
where it was predicted that the transverse current has a term
proportional to the cosine of the difference of phases between the
CDW chains. In the case when the CDW slides along one chain but is
pinned along neighbouring chains, or if the CDW moves with different
velocities in different chains, or if the CDW is pinned but  phase
slippage takes place, then the CDW phase varies with time and
alternating tunneling current is generated transversely to the chain
direction with a frequency depending on the longitudinal electric
field. When an external alternating signal acts on the sample, a
resonance can be observed at a fixed $V_{tr}$ if the frequencies of
the external and characteristic oscillations coincide. As a result,
current Shapiro steps should appear in transverse IVc that was
really observed in experiment\cite{Sinchenko12a}. In real samples
the appearance of an electric potential normal to the transport
current exists always and has been attributed to defects and
fluctuations of the critical CDW parameters such as threshold
electric field or Peierls transition temperature.\cite{Sinch11}. It
is then natural to assume that synchronisation of the generated
frequencies in NbSe$_3$ will be strongly dependent on the state of
all types of chains. Thus, sliding of high-T CDW strongly modifies
the transverse current distribution in the sample, and as a result
may destroy the synchronisation frequencies resulting for low-T CDW
sliding. However, to clarify the physical mechanism of this effect
new theoretical and experimental investigations need to be
performed.

In conclusion, we clearly observed for the first time the nonlinear
conductivity in current-voltage characteristics of NbSe$_3$,
corresponding to simultaneous sliding of both, the high-T and the
low-T, charge-density waves. We show that the interaction between
CDWs leads to strong pinning of the high-T CDW and resulting to
$4-5$ times increase of threshold electric field. In the temperature
range up to 30 K above the second Peierls transition the sliding of
the high-T CDW exhibits two thresholds, most probably because of the
local fluctuations of the low-T CDW. The superposition of an rf
current on the dc current leads to appearance of Shapiro steps on dc
IV curves for low-T CDW at $E<E_{t1}^*$ and for high-T CDW at
$E>E_{t1}^*$. Both types of Shapiro steps have never been observed
simultaneously.

\acknowledgements

The authors are thankful S.V. Zaitsev-Zotov and S.G. Zybtsev for
helpful discussions of the experimental results. The work has been
supported by Russian State Fund for the Basic Research (No.
11-02-01379-à), and partially performed in the frame of the CNRS-RAS
Associated International Laboratory between CRTBT and IRE "Physical
properties of coherent electronic states in coherent matter". The
support of ANR-07-BLAN-0136 is also acknowledged.


\begin{thebibliography}{9}

\bibitem{ECRYS11}
\textit {Electronic Crystals 2011} edited by S. Brazovskii, P.
Monceau and N. Kirova, (Physica B, vol~407, Issues 11, 2012).

\bibitem{Gruner}
G. Gr{\"u}ner, \textit{Density Waves in Solids} (Addison -- Wesley,
Reading, Massachusetts, 1994)

\bibitem{Shima}
Shima N and Kamimura H 1985 {\it Theoretical Aspects of Band
Structure and Electronic Properties of Pseudo- One-Dimensional
Solids} (Dordrecht: Reidel)

\bibitem{SinchenkoJPCM09}
A.A. Sinchenko, R.V. Chernikov, A.A. Ivanov, P. Monceau, Th. Crozes
and S.A. Brazovskii, J. Phys.: Condens. Matter {\bf 21}, (2009)
435601.

\bibitem{Gorkov89}
L. Gor'kov  and G. Gr\"uner  \textit {Charge Density Waves in
Solids} (Amsterdam: Elsevier Science, 1989)

\bibitem{Monceau12}
P. Monceau, Advances in Physics {\bf 61}, (2012) 325–581

\bibitem{Fleming83}
R.M. Fleming, C.H. Chen and D.E. Moncton, J. Physique (France) {\bf
44}, (1983) C3-1651.

\bibitem{Orlov06}
A.P. Orlov, Yu.I. Latyshev, A.M. Smolovich and P. Monceau, JETP
Lett. {\bf 84}, 89, 2006.

\bibitem{Gill80}
J.C. Gill, J. Phys. F {\bf 10}, L81 (1980).

\bibitem{Richard80}
J. Richard and P. Monceau, Solid State Commun. {\bf 33}, 635 (1980).

\bibitem{Fleming80}
R.M. Fleming, Phys. Rev. B {\bf 22}, 5606 (1980).

\bibitem{Ayari04}
A. Ayari, R. Danneau, H. Requardt, Ortega, J. E. Lorenzo, P.
Monceau, R. Currat, S. Brazovskii, and G. Gru¨bel, Phys.Rev. Lett.
{\bf 93}, (2004) 106404.

\bibitem{Noh01}
Y. Li, D. Y. Noh, J. H. Price, K. L. Ringland, J. D. Brock, S. G.
Lemay, K. Cicak, R. E. Thorne, Mark Sutton, Phys. Rev. B {\bf 63},
041103(R) (2001).

\bibitem{Sinchenko12a}
A.A. Sinchenko, P. Monceau, and T. Crozes, Phys. Rev. Lett. {\bf
108}, 046402 (2012).


\bibitem{McCarten92}
J. McCarten, D.A. DiCarlo, M.P. Maher, T.L. Adelman, and R.E.
Thorne, Phys. Rev. B {\bf 46}, 4456 (1992).

\bibitem{Zettl84}
A. Zettl and G. Gr{\"u}ner, Phys. Rev. B {\bf 29}, 755 (1984).

\bibitem{Richard82}
J. Richard, P. Monceau and M. Renard, Phys. Rev. B {\bf 25}, 948
(1992).

\bibitem{Brun10}
Ch. Brun, Zhao-Zhong Wang, P. Monceau, and S. Brazovskii, Phys. Rev.
Lett. {\bf 104}, 256403 (2010).

\bibitem{ArtemenkoVolkov84}
S.N. Artemenko and A.F. Volkov, Sov. Phys. JETP {\bf 60}, (1984),
395; JETP Lett., {\bf 83}, (1983), 368.

\bibitem{Artemenko97}
S.N. Artemenko, JETP {\bf 84}, (1997), 823.

\bibitem{Sinch11}
A.A. Sinchenko, P. Monceau and T. Crozes, JETP Lett. {\bf 93},
(2011), 56.



\end{thebibliography}
\end{document}